\documentclass{emulateapj}
\usepackage{apjfonts}
\usepackage{graphicx}

\shortauthors{Winn et al.\ 2010}
\shorttitle{RM effect for HAT-P-4 and HAT-P-14}

\begin{document}

%
\def\ltsima{$\; \buildrel < \over \sim \;$}
\def\lsim{\lower.5ex\hbox{\ltsima}}
\def\gtsima{$\; \buildrel > \over \sim \;$}
\def\gsim{\lower.5ex\hbox{\gtsima}}
                                                                                          
%

\bibliographystyle{apj}

\title{ Orbital Orientations of Exoplanets:
HAT-P-4\lowercase{b} is Prograde and HAT-P-14\lowercase{b} is Retrograde }

\author{
Joshua N.\ Winn\altaffilmark{1},
Andrew W.\ Howard\altaffilmark{2}, 
John Asher Johnson\altaffilmark{3},
Geoffrey W.\ Marcy\altaffilmark{2},\\
Howard Isaacson\altaffilmark{2},
Avi Shporer\altaffilmark{4,5}, 
G\'asp\'ar \'{A}.\ Bakos\altaffilmark{6},   
Joel D.\ Hartman\altaffilmark{6}, \\
Matthew J.\ Holman\altaffilmark{6},
Simon Albrecht\altaffilmark{1},
Justin R.\ Crepp\altaffilmark{3},
Timothy D.\ Morton\altaffilmark{3}
}


\altaffiltext{1}{Department of Physics, and Kavli Institute for
  Astrophysics and Space Research, Massachusetts Institute of
  Technology, Cambridge, MA 02139}

\altaffiltext{2}{Department of Astronomy, University of California,
  Mail Code 3411, Berkeley, CA 94720}

\altaffiltext{3}{Department of Astrophysics, and NASA Exoplanet
  Science Institute, California Institute of Technology, MC~249-17,
  Pasadena, CA 91125}

\altaffiltext{4}{Las Cumbres Observatory Global Telescope Network,
 6740 Cortona Drive, Suite 102, Santa Barbara, CA 93117}

\altaffiltext{5}{Department of Physics, Broida Hall, University of
  California, Santa Barbara, CA 93106}

\altaffiltext{6}{Harvard-Smithsonian Center for Astrophysics, 60
 Garden St., Cambridge, MA 02138}

\begin{abstract}

  We present observations of the Rossiter-McLaughlin effect for two
  exoplanetary systems, revealing the orientations of their orbits
  relative to the rotation axes of their parent stars. HAT-P-4b is
  prograde, with a sky-projected spin-orbit angle of $\lambda=-4.9\pm
  11.9$~degrees. In contrast, HAT-P-14b is retrograde, with $\lambda =
  189.1\pm 5.1$~degrees. These results conform with a previously noted
  pattern among the stellar hosts of close-in giant planets: hotter
  stars have a wide range of obliquities and cooler stars have low
  obliquities. This, in turn, suggests that three-body dynamics and
  tidal dissipation are responsible for the short-period orbits of
  many exoplanets. In addition, our data revealed a third body in the
  HAT-P-4 system, which could be a second planet or a companion star.

\end{abstract}

\keywords{planetary systems --- planets and satellites: formation ---
 planet-star interactions --- stars: rotation }

\section{Introduction}
\label{sec:introduction}

The Rossiter-McLaughlin (RM) effect, a spectroscopic phenomenon that
occurs during stellar eclipses, has recently been used to study
spin-orbit alignment for transiting exoplanets. Although the first 9
published results suggested that the orbits of close-in planets are
all well-aligned with the equatorial planes of their parent stars
(Fabrycky \& Winn 2009), the next 20 results were more diverse,
including orbits highly inclined with respect to the star's equatorial
plane (see, e.g., H\'{e}brard et al.~2008, Winn et al.~2009a, Johnson
et al.~2009) and even retrograde orbits (Anderson et al.~2010, Narita
et al.~2009, Winn et al.\ 2009b; Triaud et al.\ 2010).

These results have been marshalled as evidence against the standard
scenario for planet migration, in which disk-planet tidal interactions
cause the planet to spiral inward. Instead the results suggest that
many close-in giant planets arrived at their current locations through
gravitational perturbations from other massive bodies, followed by
tidal dissipation (Triaud et al.~2010, Winn et al.~2010a, Matsumura et
al.~2010). Another possibility is that protoplanetary disks are
frequently misaligned with the rotation of their host stars (Bate et
al.~2010, Lai et al.\ 2010).

Recently, a possible trend emerged from the results: misaligned
systems tend to have stars with effective temperatures exceeding about
6250~K, or masses $\gsim$1.2~$M_\odot$. The evidence for this pattern
is based not only on RM observations (Winn et al.~2010a) but also on
the line-of-sight stellar rotation velocities of transit hosts
(Schlaufman 2010). This trend may indicate that planet formation and
migration are fundamentally different for low-mass stars than for
high-mass stars, for which there is already evidence in the
distributions of planet mass and period (Bowler et al.~2010). Another
possibility is that the formation and migration processes are similar,
but that the subsequent tidal evolution is different (Winn et
al.~2010a). In this hypothesis, cool stars are observed to have low
obliquities because tidal evolution drove them into alignment, while
hot stars retain their ``primordial'' obliquities because of their
thinner (or absent) outer convection zones and consequently slower
rates of tidal dissipation.

Although the trend seems clear, it is difficult to assess its true
significance because many possible variables were examined before
alighting on stellar temperature and mass. The only way to be sure is
to gather more data. This paper presents results for the next two
systems we observed after the trend had been identified. Both systems
have short-period giant planets, but HAT-P-4 is ``cool'' ($T_{\rm eff}
= 5860\pm 80$~K; Kov\'{a}cs et al.~2007) while HAT-P-14 is ``hot''
($6600\pm 90$~K; Torres et al.~2010).  We present the observations of
these systems in \S~2, the analysis and results in \S~3, and a
discussion in \S~4.

\section{Observations}
\label{sec:observations}

Our spectroscopic observations employed the High Resolution
Spectrograph (HIRES; Vogt et al.~1994) of the Keck~I 10m telescope, on
Mauna Kea, Hawaii. We gathered 35 spectra of HAT-P-4 on the night of
2010~March~29/30, and 44 spectra of HAT-P-14 on the night of
2010~April~27/28, in both cases spanning a predicted transit of the
planet. An additional 14 spectra of HAT-P-4 were gathered on other
nights, at essentially random orbital phases.

We used the standard instrument settings and observing procedures of
the California Planet Search (Howard et al.~2009). The iodine gas
absorption cell was used to track the instrumental response and
wavelength scale. The relative radial velocity (RV) of each spectrum
was measured with respect to an iodine-free template spectrum, using a
descendant of the algorithm of Butler et al.~(2006). Measurement
errors were estimated from the scatter among the fits to individual
spectral segments spanning a few Angstroms. Tables~\ref{tbl:rv-hat14}
and \ref{tbl:rv-hat4} give the RVs, including re-reductions of data
presented earlier by Kov\'acs et al.~(2007) and Torres et al.~(2010).

We also conducted photometric observations of HAT-P-4, in order to
refine the transit ephemeris and other system parameters. We observed
the transit of 2010~March~29/30 (the same night as the Keck
observations) with the Faulkes Telescope North (FTN) 2m telescope, on
Mauna Haleakala, Hawaii. We used the Spectral Instruments camera with
an SDSS $i$ filter and $2\times 2$ binning, giving a pixel scale of
$0\farcs304$ and a $10\farcm5$ field of view. Unfortunately the guider
malfunctioned that night. The transit of 2010~May~7/8 was observed
with the Fred L.\ Whipple Observatory (FLWO) 1.2m telescope on Mt.\
Hopkins, Arizona. We used KeplerCam with an SDSS $i$ filter and
$2\times 2$ binning, giving a pixel scale of $0\farcs67$ and a
$23\farcm1$ field of view. After standard debiasing and flat-fielding
operations, we performed differential aperture photometry of HAT-P-4
and several other stars in the field.

\bigskip
\smallskip

\section{Analysis}
\label{sec:analysis}

\smallskip
\smallskip

\subsection{HAT-P-14}

\smallskip
 
We begin with HAT-P-14, for which the analysis proved simpler. The
model for the RV data took the form
\begin{equation}
V_{{\rm calc}}(t) = V_{\rm orb}(t) + V_{\rm RM}(t) + \gamma,
\end{equation}
where $V_{\rm orb}(t)$ is the radial component of a Keplerian orbit,
$V_{\rm RM}(t)$ is the anomalous velocity due to the RM effect, and
$\gamma$ is an arbitrary offset related to the barycentric RV of the
template spectrum. To model the RM effect, we used the technique of
Winn et al.~(2005), which entails the construction and Doppler
analysis of simulated spectra exhibiting the RM effect. The resulting
formula for the anomalous velocity was
\begin{equation}
V_{\rm RM}(t) = \Delta f(t)~v_p(t)
\left[1.58 - 0.883\left(\frac{v_p(t)}{8.4~{\rm km~s}^{-1}}\right)^2\right],
\end{equation}
where $\Delta f$ is the fractional loss of light during the transit,
$v_p$ is subplanet velocity (defined as the line-of-sight component of
the stellar rotation velocity at the position of the photosphere
directly behind the center of the planet), and the figure of
8.4~km~s$^{-1}$ is the value of $v\sin i_\star$ estimated by Torres et
al.~(2010). In calculating $\Delta f$, we adopted a linear
limb-darkening law with a coefficient of 0.6288, based on
interpolation of the tables of Claret~(2004). In calculating $v_p(t)$,
we neglected differential rotation, and allowed the stellar rotation
axis and the orbit normal to be separated by an angle $\lambda$ on the
sky plane. For a diagram of the coordinate system, see Ohta et
al.~(2005) or Fabrycky \& Winn (2009).

Many of the parameters of the Keplerian orbit and of the eclipses have
been tightly constrained by previous observations. We adopted priors
on those parameters, to keep the number of free parameters in our
model to a minimum. Specifically, we used a fitting statistic
\begin{eqnarray}
\label{eq:chi2-hat14}
\chi^2 & = & \sum_{i=1}^{44}
\left[ \frac{ V_{\rm obs}(t_i) - V_{\rm calc}(t_i) }
  {\sigma_V} \right]^2 + \;\;\;\nonumber \\
&  &   \left(\frac{T_c - 2\,454\,875.28938}{0.00047}\right)^2 + 
     \left(\frac{P_{\rm days} - 4.627669}{0.000005}\right)^2 + \;\;\;\nonumber \\
&  &   \left(\frac{T_{\rm days} - 0.0912}{0.0017}\right)^2 + 
     \left(\frac{\tau_{\rm days} - 0.0287}{0.0026}\right)^2 + \;\;\;\nonumber \\
&  &   \left(\frac{R_p/R_\star - 0.0805}{0.0015}\right)^2 + 
     \left(\frac{K_\star - 219.0~{\rm m~s}^{-1}}{3.3~{\rm m~s}^{-1}}\right)^2 + \;\;\;\nonumber \\
&  &   \left(\frac{v\sin i_\star - 8.4~{\rm km~s}^{-1}}{0.50~{\rm km~s}^{-1}}\right)^2,
\end{eqnarray}
where the first term is the usual sum of squared residuals, and the
other terms enforce Gaussian priors. In this expression, $P_{\rm
  days}$ is the orbital period in days, $T_c$ is a particular time of
inferior conjunction (in the HJD$_{\rm UTC}$ system); $T_{\rm days}$
is the time between first and fourth contact; $\tau_{\rm days}$ is the
time between first and second contact; $R_p/R_\star$ is the
planet-to-star radius ratio; $K_\star$ is the radial velocity
semiamplitude of the star's Keplerian orbit; and $v\sin i_\star$ is
the line-of-sight component of the star's equatorial rotation
velocity. All the numerical values and uncertainties are taken from
Torres et al.~(2010). We held constant the orbital eccentricity
$e=0.107$ and argument of pericenter $\omega=94^\circ$, since those
parameters have little effect on the model once the transit ephemeris
is specified.

For $\sigma_V$, we used the quadrature sum of the internally-estimated
measurement error (typically 4~m~s$^{-1}$; see Table 1) and a
``jitter'' term of 7.8~m~s$^{-1}$, a value determined from the
condition $\chi^2 = N_{\rm dof}$. Thus, the excess noise was assumed
to be uncorrelated in time. It is comparable in magnitude to the
jitter term of 7.3~m~s$^{-1}$ used by Torres et al.~(2010).

\begin{figure*}[ht]
\begin{center}
\leavevmode
\hbox{
\epsfxsize=7.5in
\epsffile{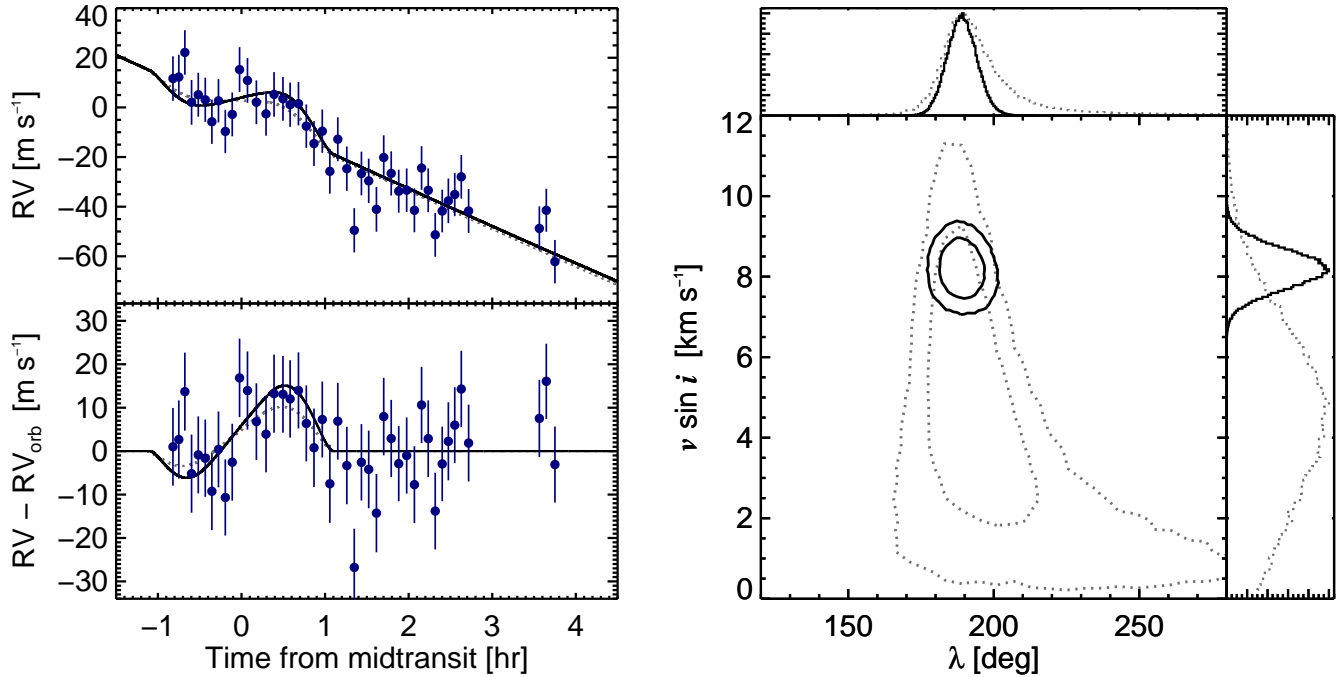}}
\end{center}
\vspace{-0.1in}
\caption{{\bf Results for HAT-P-14.}
{\it Left.}---Apparent radial velocity variation
on the night of 2010~April~27/28, spanning a transit.
The top panel shows the observed RVs.
For the bottom panel, the best-fitting orbital
model was subtracted, thereby isolating the anomalous
RV due to the RM effect. The black curve shows the best-fitting model
with a prior constraint $v\sin i_\star = 8.4\pm 0.5$~km~s$^{-1}$,
and the dotted curve is the best-fitting model with no prior constraint
on $v\sin i_\star$.
{\it Right.}---Joint constraints on $\lambda$ and $v\sin i_\star$.
The contours represent 68.3\% and 95.4\%
confidence limits. The
marginalized posterior probability distributions
are shown on the sides of the contour plot. The solid and dotted curves
show the results with and without the prior constraint on $v\sin
i_\star$.
\vspace{0.25in}
\label{fig:hat14}}
\end{figure*}

There were only two completely free parameters in our model: $\gamma$,
the overall RV offset; and $\lambda$, the projected spin-orbit angle.
Parameter optimization and error estimation were achieved with a
Markov Chain Monte Carlo (MCMC) algorithm, using Gibbs sampling and
Metropolis-Hastings stepping. Table~\ref{tbl:params-hat14} summarizes
the results. The first section lists all of the adjustable model
parameters, for which uniform priors were adopted. The last section
gives some results for other quantities that were computed based on
the model parameters, or taken from Torres et al.~(2010). The quoted
values and ranges are based on the 50\%, 15.85\%, and 84.15\%
confidence levels of the marginalized
posteriors. Figure~\ref{fig:hat14} shows the RV data and the results
for $v\sin i_\star$ and $\lambda$.

The result for $\lambda$ is $189.1\pm 5.1$~degrees, indicating that
the directions of orbital motion and stellar rotation are nearly
opposite as projected on the sky. This result could be anticipated
from a visual inspection of Figure~\ref{fig:hat14}, which shows that
the anomalous RV was a redshift in the second half of the transit and
(less obviously) a blueshift in the first half. This is an inversion
of the more familiar pattern of a well-aligned system. The orbit of
HAT-P-14b is strongly misaligned with the rotational plane of its
parent star.

Although the finding of a retrograde orbit is robust, the small
uncertainty in $\lambda$ depends critically on the prior constraint on
$v\sin i_\star$. If that constraint is dropped, a slower rotation rate
is favored ($v\sin i_\star = 4.5\pm 2.4$~km~s$^{-1}$), allowing a
broader range of spin-orbit angles ($\lambda =
192.0_{-8.7}^{+15.6}$~degrees). These results are also illustrated by
the dotted lines in Figure~\ref{fig:hat14}. One possible reason for
the smaller result for $v\sin i_\star$ is differential rotation: the
RM effect depends on the rotation rate over the range of latitudes
spanned by the transit chord, which may differ from the
spectroscopically-estimated equatorial rotation rate. The near-grazing
transit of HAT-P-14, in particular, may produce the most extreme
possible difference. Further observations with a higher
signal-to-noise ratio might be able to identify the specific signal of
differential rotation (Gaudi \& Winn 2007).

\subsection{HAT-P-4}

\bigskip

\begin{figure}[ht]
\epsscale{1.1}
\plotone{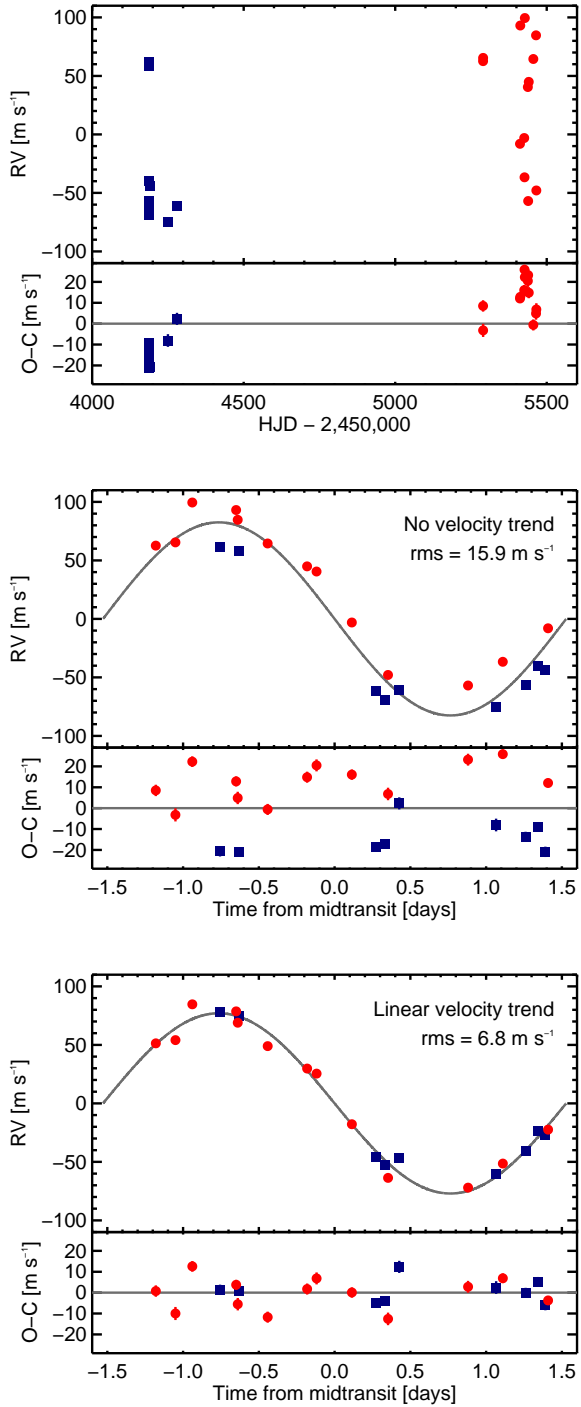}
\caption{{\bf Evidence for a third body in the HAT-P-4 system.}
{\it Top.}---Relative RV as a function of time. Squares represent
data from 2007, and circles represent data from 2010.
Below the data are the residuals between the data and the best-fitting
model involving a single planet on a circular orbit.
{\it Middle.}---Same, but plotted as a function of the orbital phase
of the planet, in days. The residuals from 2010 are systematically
higher than those from 2007, which is evidence for an excess
radial acceleration and hence an additional gravitating
body in the HAT-P-4 system.
{\it Bottom.}---Same as the middle panel, but for a model
which also includes a free parameter $\dot{\gamma}$ representing a constant
acceleration. The result was $\dot{\gamma}= 0.0246\pm
0.0026$~m~s$^{-1}$~day$^{-1}$.
\label{fig:hat4_orb}}
\end{figure}

The case of HAT-P-4 was more complicated, partly because our RV
observations revealed a third body in the system.
Figure~\ref{fig:hat4_orb} illustrates that a single planet on a
circular orbit no longer provides a satisfactory description of the
data. The best-fitting model has $\chi^2 = 925.7$ with 18 degrees of
freedom. The residuals have an rms of 15.9~m~s$^{-1}$ and are highly
correlated, with almost all of the most recent RVs lying above the
model curve. Allowing the orbit to be eccentric reduces $\chi^2$ to
906.0, but the residuals still have an rms of 15~m~s$^{-1}$ and show
the same pattern.

A much better fit is obtained when the circular orbit is supplemented
by a constant acceleration $\dot{\gamma}$. This model, illustrated in
the lower panel of Figure~\ref{fig:hat4_orb}, gives $\chi^2=137.0$ and
more randomly scattered residuals with an rms residual of
6.8~m~s$^{-1}$. An MCMC analysis gives $\dot{\gamma}= 0.0246\pm
0.0026$~m~s$^{-1}$~day$^{-1}$.

The constant acceleration may represent the RV variation due to a
companion star or planet whose orbit is longer than the three-year
span of our observations. Assuming it is a low-mass body on a nearly
circular orbit, we may set $\dot{\gamma} \sim GM_c \sin i_c/a_c^2$,
giving an order-of-magnitude constraint
\begin{equation}
\left( \frac{M_c \sin i_c}{M_{\rm Jup}} \right)
\left( \frac{a_c}{{\rm 10~AU}} \right)^{-2} \sim 5.03\pm 0.53.
\end{equation}
However, given the limited time sampling of our data it is also
possible that the companion has a shorter period. We have not found
any compelling two-planet models, but we are continuing to gather
additional RV data and will report elsewhere on the results. The
pertinent conclusions for this study are (1) there is a third body in
the system, and (2) the mass and orbital parameters of HAT-P-4b are
subject to systematic errors due to the unknown influence of the third
body on the RV data. For the latter reason, in our analysis of the
transit data we did not employ a prior constraint on $K_\star$, as we
did for HAT-P-14.

\begin{figure}[ht]
\epsscale{1.0}
\plotone{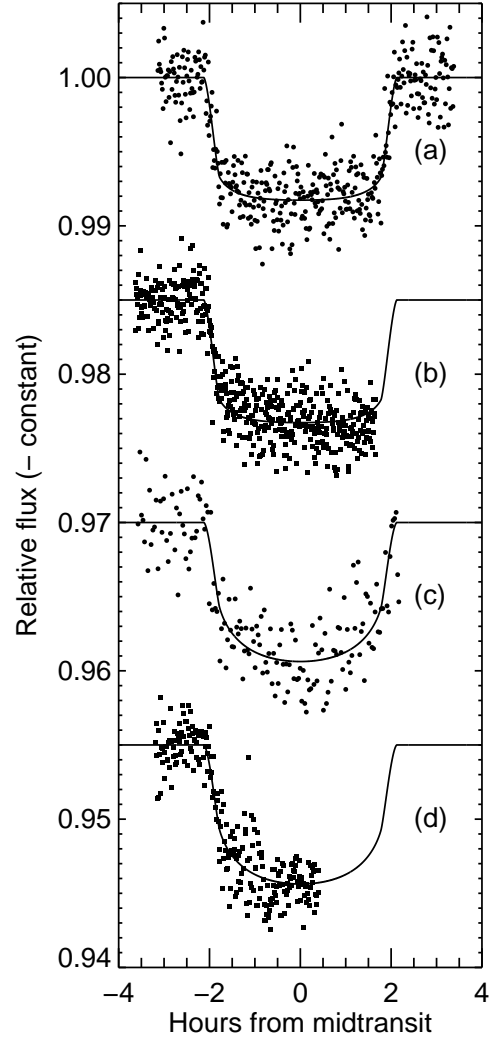}
\caption{{\bf Transit light curves of HAT-P-4.}
(a) and (b): FLWO $z$-band data, from Kov\'acs et al.~(2007).
(c): FTN $i$-band data, from our new observations. These data
were obtained simultaneously with the Keck RM observations.
(d): FLWO $i$-band data, from our new observations.
The best-fitting model light curves are overplotted.
\label{fig:hat4_newlc}}
\end{figure}

Another difference in our analysis is that we fitted the available
photometric data along with the transit-night RV data, because the new
data offer better constraints on the transit ephemeris, depth,
duration, and partial duration. Our photometric model was taken from
the Transit Light Curve project (see, e.g., Holman et al.~2006, Winn
et al.~2009c). In brief, we used the Mandel \& Agol (2002) formulas
for a quadratic limb-darkening law, as implemented by P\'al
(2008). The linear coefficient was allowed to vary freely, and the
quadratic coefficient was held fixed at the Claret~(2004) value
(0.3418 for $i$-band, and 0.3395 for $z$-band). The out-of-transit
magnitude was allowed to be a linear function of airmass, to account
for color-dependent differential extinction. The errors for each light
curve were set equal to $\beta\sigma_1$, where $\sigma_1$ is the
root-mean-squared (rms) residual, and $\beta$ accounts for time
correlations, using the method of Pont et al.\ (2006) as implemented
by Winn et al.\ (2009c). Averaging times of 10-30 minutes were used to
compute $\beta$, giving results (in chronological order) of 1.01,
1.73, 1.69, and 1.01. Parameter estimation was performed by the MCMC
method. The photometric data and the best-fitting models are plotted
in Figure~\ref{fig:hat4_newlc}.

The formula for the anomalous velocity was
\begin{equation}
V_{\rm RM}(t) = \Delta f(t)~v_p(t)
\left[1.36 - 0.628\left(\frac{v_p(t)}{5.5~{\rm km~s}^{-1}}\right)^2\right],
\end{equation}
and the fitting statistic was
\begin{eqnarray}
\chi^2 & = &
\sum_{i=1}^{35}
\left[ \frac{ V_{\rm obs}(t_i) - V_{\rm calc}(t_i) } {\sigma_V}
\right]^2 +
\sum_{i=1}^{1798}
\left[ \frac{ f_{\rm obs}(t_i) - f_{\rm calc}(t_i) } {\sigma_f}
\right]^2 + \nonumber \\
 &  &  \left(\frac{v\sin i_\star - 5.50~{\rm km~s}^{-1}}{0.55~{\rm
           km~s}^{-1}}\right)^2, \;\;\;
\end{eqnarray}
using terminology similar to that of Eqn.~(\ref{eq:chi2-hat14}). The
only free parameters that were wholly dependent on the RV data were
$\gamma$, $\lambda$, and $K_\star$. The jitter term was $\sigma_V =
5.2$~m~s$^{-1}$.

\begin{figure*}[ht]
\begin{center}
\leavevmode
\hbox{
\epsfxsize=7.5in
\epsffile{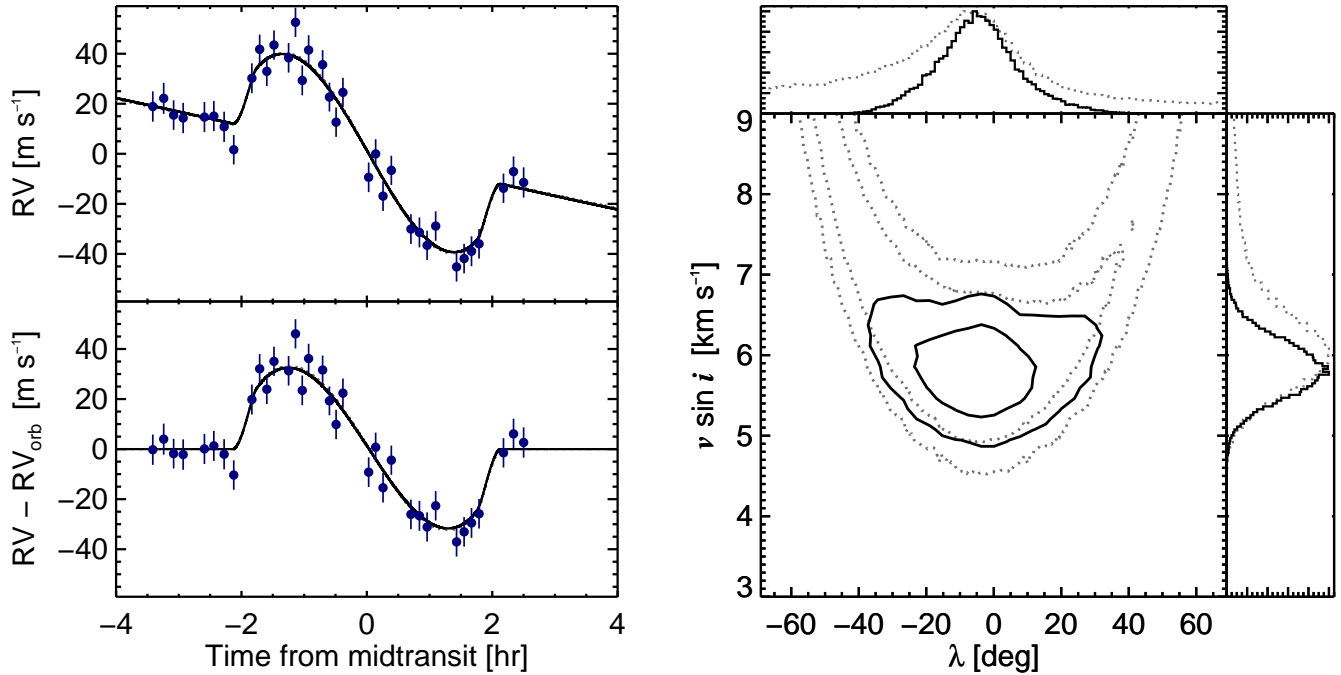}}
\end{center}
\vspace{-0.2in}
\caption{{\bf Results for HAT-P-4.}
{\it Left.}---Observed radial velocity variation
on the night of 2010~March~29/30, spanning a transit.
The top panel shows the observed RVs.
For the bottom panel, the best-fitting orbital
model was subtracted, thereby isolating the anomalous
RV due to the RM effect. The solid black curve shows the best-fitting model
with a prior constraint $v\sin i_\star = 5.5\pm 0.5$~km~s$^{-1}$.
The best-fitting model with no prior constraint
on $v\sin i_\star$ is plotted with a dotted curve, although it is hard to
distinguish from the solid curve.
{\it Right.}---Joint constraints on $\lambda$ and $v\sin i_\star$.
The contours represent 68.3\% and 95.4\%
confidence limits. The
marginalized posterior probability distributions
are shown on the sides of the contour plot.
The solid and dotted curves
show the results with and without the prior constraint on $v\sin i_\star$.
\label{fig:hat4}}
\end{figure*}

Table~\ref{tbl:params-hat4} gives the results, based on the 15.85\%,
50\%, and 84.15\% confidence levels of the marginalized {\it a
  posteriori} distributions. Figure~\ref{fig:hat4} shows the RV data,
and the posteriors for the RM parameters $v\sin i_\star$ and
$\lambda$. In particular, the projected spin-orbit angle is
$\lambda=-4.9\pm 11.9$~degrees, consistent with good alignment between
the rotational and orbital angular momentum.

As before, the quantitative results hinge on the prior constraint on
$v\sin i_\star$. By repeating the analysis with no such constraint, we
find $v\sin i_\star = 6.4_{-0.7}^{+4.7}$~km~s$^{-1}$ and $\lambda =
-11_{-39}^{+29}$~degrees. These much broader results are also
illustrated by the gray lines in Figure~\ref{fig:hat4}. The only
well-constrained combination of those two parameters is $v\sin
i_\star\cos\lambda = 5.77\pm 0.41$~km~s$^{-1}$.

\bigskip

\section{Discussion}
\label{sec:explanation}

It might seem surprising that tighter bounds on $\lambda$ were
obtained for HAT-P-14 than for HAT-P-4, given that the signal-to-noise
ratio of the RM effect is higher for HAT-P-4. This is a consequence of
the difference in the impact parameter (the minimum sky-projected
distance between the planet and the star, in units of the stellar
radius). The interpretation of the RM signal is most robust for
systems with a high impact parameter, because in such cases the two
key parameters $v\sin i_\star$ and $\lambda$ play distinct roles:
$v\sin i_\star$ controls the amplitude of the signal and $\lambda$
controls its shape (i.e.\ the phase of the transit when the anomalous
RV switches from positive to negative, or vice versa). For systems
with low impact parameters, such as HAT-P-4, the shape of the RM
signal is nearly independent of $\lambda$ and both parameters control
the amplitude. This leads to a strong degeneracy between those two
parameters (Gaudi \& Winn 2007).

Therefore, an external constraint on $v\sin i_\star$ is crucial for
the determination of $\lambda$ in systems with low impact parameters.
In our study we have used a prior based on the line broadening
observed in the star's optical spectrum. However, it must be
acknowledged that the resulting estimate of $v\sin i_\star$ is subject
to systematic error due to uncertainties in the competing effects of
macroturbulence and other broadening mechanisms. This is especially
problematic for cool, low-mass stars for which turbulent and
instrumental broadening exceed rotational broadening; examples
are TrES-1 (Narita et al.~2007) and WASP-4 (Triaud et
al.~2010). For HAT-P-4 the situation is better because rotational
broadening is expected to be at least as important as turbulent
broadening.\footnote{Given HAT-P-4's approximate spectral type of
  F7/G0, the expected macroturbulent velocity is
  $\approx$4.5~m~s$^{-1}$ (Gray~2008, p.\ 443) as compared to the
  inferred $v\sin i_\star$ of $5.5\pm 0.5$~m~s$^{-1}$.}

We turn now to the question posed at the beginning of this paper: do
hot stars have high obliquities?  Specifically, among the host stars
of close-in planets, are those with $T_{\rm eff} \lsim 6250$~K more
likely to be aligned with the planetary orbits than hotter stars? We
have found that HAT-P-4 is a well-aligned cool star ($\lambda=-4.9\pm
11.9$~deg, $T_{\rm eff} = 5860\pm 80$~K), and HAT-P-14 is a misaligned
hot star ($\lambda=189.1\pm 5.1$~deg, $T_{\rm eff} = 6600\pm
90$~K). Therefore, these new data strengthen the trend that was
observed by Winn et al.~(2010a).

Schlaufman (2010) found a similar pattern using a different technique,
involving a comparison between the observed and expected line-of-sight
rotational velocities of the stars with transiting planets. He
described the pattern in terms of stellar mass rather than effective
temperature. Indeed, those two parameters are strongly correlated for
dwarf stars, with scatter due to metallicity and age. For HAT-P-4 and
HAT-P-14, the stellar masses are $1.26_{-0.14}^{+0.06}~M_\odot$ and
$1.386\pm 0.045~M_\odot$ respectively. For these systems there is a
clearer contrast in effective temperature (6.1$\sigma$) than mass
(1.7$\sigma$). The very different results for $\lambda$ suggest that
effective temperature is more closely related to obliquity than
mass. This in turn would support the hypothesis of Winn et al.~(2010a)
that the differing obliquities are a consequence of differing internal
structure of the stars, and specifically the depth of the outer
convective zone, since this structural difference is more closely
related to effective temperature than mass (Pinsonneault et al.~2001).

It will be interesting to examine all the systems for which the RM
effect has been measured, to see whether temperature or mass is more
important, and whether there are other variables related to
obliquity. We defer such a study for the future, to allow the sample
size to grow substantially since the last such analysis by Winn et
al.~(2010a). One variable that will be especially interesting to
assess is the presence or absence of a third body. A migration
mechanism involving the Kozai effect, which has been invoked to
explain high obliquities, requires the existence of a third body (see,
e.g., Fabrycky \& Tremaine 2007). In planet-planet scattering
scenarios, a third body may also be present, although it could have
been ejected (see, e.g., Chatterjee et al.~2008). Evidence for
additional bodies has been found in some well-aligned systems such as
HAT-P-13 (Bakos et al.~2010, Winn et al.~2010b) and HAT-P-4 (this
study), along with some misaligned systems such as HAT-P-7 (Winn et
al.~2009a), HD~80606 (Naef et al.~2001), HAT-P-11 (Bakos et al.~2010,
Winn et al.~2010c), and WASP-8 (Queloz et al.~2010). A systematic
multiplicity study would be illuminating (see, e.g., Narita et
al.~2010), as would a comparison between the observed obliquity
distribution and that predicted by the Kozai model (Fabrycky \& Winn
2009; Morton \& Johnson 2010).

It will also be interesting to extend these studies to
multi-transiting systems such as Kepler-9 (Holman et al.~2010). This
will allow the mutual inclinations of the orbits to be determined,
along with the obliquity of the star (Fabrycky 2009, Ragozzine \&
Holman 2010). If the mechanism that causes spin-orbit misalignments is
related to star formation (Bate et al.~2010) or star-disk interactions
(Lai et al.~2010), and is not related to the planet, then one would
expect the planetary orbits to be well-aligned and the star to be
tipped away from their common orbital plane. In contrast, if the
close-in planet ``pile-up'' has an origin in dynamical scattering and
tidal dissipation, then the planetary orbits would be highly inclined.

\acknowledgments We thank Teruyuki Hirano for interesting discussions
related to this work, Amaury Triaud and the anonymous referee for
insightful comments on the manuscript, and John Southworth for making
available his useful code {\sc jktld} for calculating theoretical limb
darkening coefficients. We gratefully acknowledge support from the
NASA Origins program through awards NNX09AD36G and NNX09AB33G, and the
MIT Class of 1942. S.A.\ acknowledges support from a NWO Rubicon
fellowship.

This paper uses observations obtained with facilities of the Las
Cumbres Observatory Global Telescope. Some data presented herein were
obtained at the W.M.~Keck Observatory, which is operated as a
scientific partnership among the California Institute of Technology,
the University of California, and the National Aeronautics and Space
Administration, and was made possible by the generous financial
support of the W.M.~Keck Foundation. We extend special thanks to those
of Hawaiian ancestry on whose sacred mountain of Mauna Kea we are
privileged to be guests.  Without their generous hospitality, the Keck
observations presented herein would not have been possible.

\eject

\begin{deluxetable}{lcc}

\tabletypesize{\scriptsize}
\tablecaption{Relative Radial Velocity Measurements of HAT-P-14\label{tbl:rv-hat14}}
\tablewidth{0pt}

\tablehead{
\colhead{HJD$_{\rm UTC}$} &
\colhead{RV [m~s$^{-1}$]} &
\colhead{Error [m~s$^{-1}$]}
}

\startdata
  $  2454602.85804$  &  $   -187.65$  &  $   3.64$  \\
  $  2454603.10267$  &  $   -198.11$  &  $   3.64$  \\
  $  2454603.86302$  &  $   -142.90$  &  $   3.70$  \\
  $  2454604.09555$  &  $    -81.34$  &  $   4.05$  \\
  $  2454633.99342$  &  $    212.00$  &  $   4.03$  \\
  $  2454634.93451$  &  $    -93.25$  &  $   3.85$
\enddata

\tablecomments{The RV was measured relative to an arbitrary template
  spectrum; only the differences are significant. The uncertainty
  given in Column 3 is the internal error only and does not account
  for any possible ``stellar jitter.'' (We intend for this table to be
  available in its entirety in a machine-readable form in the online
  journal. A portion is shown here for guidance regarding its form and
  content.)}

\end{deluxetable}

\begin{deluxetable}{lcc}

\tabletypesize{\scriptsize}
\tablecaption{Relative Radial Velocity Measurements of HAT-P-4\label{tbl:rv-hat4}}
\tablewidth{0pt}

\tablehead{
\colhead{HJD$_{\rm UTC}$} &
\colhead{RV [m~s$^{-1}$]} &
\colhead{Error [m~s$^{-1}$]}
}

\startdata
  $  2454186.98523$  &  $     55.67$  &  $   2.44$  \\
  $  2454187.11242$  &  $     52.15$  &  $   2.12$  \\
  $  2454188.01161$  &  $    -68.10$  &  $   2.15$  \\
  $  2454188.07151$  &  $    -75.14$  &  $   1.92$  \\
  $  2454189.00175$  &  $    -62.98$  &  $   2.36$  \\
  $  2454189.08264$  &  $    -46.17$  &  $   2.18$
\enddata

\tablecomments{The RV was measured relative to an arbitrary template
  spectrum; only the differences are significant. The uncertainty
  given in Column 3 is the internal error only and does not account
  for any possible ``stellar jitter.'' (We intend for this table to be
  available in its entirety in a machine-readable form in the online
  journal. A portion is shown here for guidance regarding its form and
  content.)}

\end{deluxetable}

\normalsize

\begin{deluxetable}{lc}
 
\tabletypesize{\normalsize}
\tablecaption{Parameters for HAT-P-14\label{tbl:params-hat14}}
\tablewidth{0pt}
 
\tablehead{
\colhead{Parameter} &
\colhead{Value}
}

\startdata
{\it Model parameters} \\[0.01in]
\hline
Projected spin-orbit angle, $\lambda$~[deg]                     & $189.1\pm 5.1$ \\
Projected stellar rotation rate, $v \sin i_\star$~[km~s$^{-1}$]    & $8.18\pm 0.49$ \\   
RV offset [m~s$^{-1}$]                                           & $20.6 \pm 1.9$ \\ 
Velocity semiamplitude, $K_\star$~[m~s$^{-1}$]        & $218.9 \pm 5.7$ \\
Orbital period, $P$~[days]                         & $4.6276690 \pm 0.0000050$ \\
Midtransit time~[HJD$_{\rm UTC}$]                     & $2,455,314.91794 \pm 0.00066$ \\
Planet-to-star radius ratio, $R_p/R_\star$           & $0.0800\pm 0.0015$ \\ 
Orbital inclination, $i$~[deg]                     & $83.52\pm 0.22$  \\ 
Fractional stellar radius, $R_\star/a$               & $0.1127\pm 0.0033$    \\[0.01in]
\hline
{\it Other parameters (derived from model parameters, or from elsewhere)} \\[0.01in]
\hline
Transit duration, first to fourth contact [days]   & $0.0910\pm 0.0035$ \\ 
Transit ingress or egress duration [days]          & $0.0294\pm 0.0029$ \\
Transit impact parameter                           & $0.894\pm 0.013$\\
Orbital eccentricity (Torres et al.~2010) & $0.107\pm 0.013$ \\ 
Argument of pericenter [deg] (Torres et al.~2010) & $94\pm 4$  \\
Stellar mass, $M_\star$ [$M_\odot$] (Torres et al.~2010) & $1.386 \pm 0.045$ \\
Stellar radius, $R_\star$ [$R_\odot$]  & $1.468 \pm 0.042$ \\
Planetary mass, $M_p$ [$M_{\rm Jup}$]  & $2.232 \pm 0.058$ \\ 
Planetary radius, $R_p$ [$R_{\rm Jup}$] & $1.142\pm 0.033$
\enddata

\end{deluxetable}

\begin{deluxetable}{lc}

\tabletypesize{\normalsize}
\tablecaption{Parameters for HAT-P-4\label{tbl:params-hat4}}
\tablewidth{0pt}
 
\tablehead{
\colhead{Parameter} &
\colhead{Value}
}

\startdata
{\it Model parameters} \\
\hline
Projected spin-orbit angle, $\lambda$~[deg]                     & $-4.9\pm 11.9$ \\
Projected stellar rotation rate, $v \sin i_\star$~[km~s$^{-1}$]    & $5.83\pm 0.35$ \\ 
RV offset [m~s$^{-1}$]                                           & $-5.7 \pm 1.3$ \\ 
Velocity semiamplitude, $K_\star$~[m~s$^{-1}$] & $66.9 \pm 8.1$ \\
Orbital period, $P$~[days]                         & $3.0565195 \pm 0.0000025$ \\
Midtransit time~[HJD$_{\rm UTC}$]                     & $2,455,285.03216 \pm 0.00073$ \\
Planet-to-star radius ratio, $R_p/R_\star$           & $0.08697_{-0.00045}^{+0.00052}$ \\ 
Orbital inclination, $i$~[deg]                     & $88.76_{-1.38}^{+0.89}$  \\ 
Fractional stellar radius, $R_\star/a$               & $0.1690_{-0.0051}^{+0.0064}$ \\ 
\hline
{\it Other parameters (derived from model parameters, or from elsewhere)} \\[0.01in]
\hline 
Transit duration, first to fourth contact [days]   & $0.1775_{-0.0048}^{+0.0053}$ \\ 
Transit ingress or egress duration [days]          & $0.01465_{-0.00054}^{+0.00092}$ \\
Transit impact parameter                           & $0.1284_{-0.092}^{+0.137}$ \\
Orbital eccentricity & $0$ (assumed) \\ 
Stellar mass, $M_\star$ [$M_\odot$] (Kov\'acs et al.~2007) & $1.26 \pm 0.10$ \\
Stellar radius, $R_\star$ [$R_\odot$]  & $1.617_{-0.050}^{+0.057}$ \\
Planetary mass, $M_p$ [$M_{\rm Jup}$]  & $0.556 \pm 0.068$ \\ 
Planetary radius, $R_p$ [$R_{\rm Jup}$] & $1.367_{-0.044}^{+0.052}$
\enddata

\end{deluxetable}

\end{document}